\documentclass[graybox]{svmult}

\usepackage{helvet}         
\usepackage{courier}        
\usepackage{type1cm}        
\usepackage{makeidx}         
\usepackage{graphicx}        
\usepackage{multicol}        
\usepackage[bottom]{footmisc}

\usepackage{amsmath}
\usepackage{amssymb}
\usepackage{amsfonts}
\usepackage{amscd}
\usepackage{bm}
\usepackage{stmaryrd}
\usepackage{bbold}
\usepackage{isomath}
\usepackage{xcolor}
\usepackage{theorem}
\usepackage{xspace}
\usepackage{tensor}
\input xy
\xyoption{all}

\def\J{{\cal J}}

\def\Z{\mathbb Z}
\def\R{\mathbb R}
\def\C{\mathbb C}

\def\demi{{\frac{1}{2}}}

\newcommand{\beq}{\begin{equation}}
\newcommand{\eeq}{\end{equation}}
\newcommand{\ba}{\begin{array}}
	\newcommand{\ea}{\end{array}}
\newcommand{\beqa}{\begin{eqnarray}}
\newcommand{\eeqa}{\end{eqnarray}}

\def\JJ{\mathfrak J}
\def\demi{\frac{1}{2}}

\def\JJ{\mathfrak J}
\def\j{\mathcal J}
\def\k {\mathfrak{J}}
\def\d {\mathfrak{d}}

\def\demi{\frac{1}{2}}

\begin{document}

\title*{ 
 Landau levels via Jordan superalgebras}
\author{Alessio Marrani 
and Todor Popov}
\institute{Alessio Marrani \at Centre for Mathematics and Theoretical Physics, University of Hertfordshire, UK \\
\email{a.marrani@herts.ac.uk}
\and Todor Popov \at Math and Science Department, {American University in Bulgaria, BG  } \\
{INRNE at Bulgarian Academy of Sciences, BG} \\
\email{tpopov@aubg.edu}}

\maketitle

\abstract{ The  goal of this note is to show that Jordan 
algebras and superalgebras provide an elegant and concise
language for formulating quantum mechanical problems
with  inherent 
(super)conformal symmetry.
The superconformal symmetries of 
the quantum  MICZ-Kepler model  and its dual oscillator  realization  in $\R^2$ are reviewed through the lens of the Tits-Kantor-Koecher 
correspondence:
 Kaplansky $\JJ \R^{1|2}$ and  Exceptional $\JJ F^{6|4}$ Jordan superalgebras
provide a natural framework
for reconstructing   (variously extended) 
superconformal algebras hidden in 
the  Landau levels of an electron in an external magnetic field.
}


\section{Introduction}

The Tits-Kantor-Koecher  (TKK)  construction \cite{Kantor, Koecher, Tits} assigns to a given Euclidean Jordan algebra $\JJ$ a hierarchy of Lie algebras 
embedded into the conformal algebra $\mathfrak{co}(\JJ)$,
\beq
\mathfrak{der}(\JJ) \subset\mathfrak{str}(\JJ)  \subset\mathfrak{co}(\JJ)
\label{Tkk} \ .
\eeq
Supersymmetric (i.e. superized) TKK has been investigated by G\"unaydin  \cite{G, MG} (see also \cite{BC}).
On the other hand, along the years Jordan algebras have been employed by Meng \cite{Meng} for defining  families of classical (Kepler/Coulomb)  and quantum (hydrogen-like) systems, with or without magnetic charge, and they also proved to be crucial for exhibiting the hidden conformal symmetries of the quantum Kepler models \cite{Meng} (see also \cite{DNP, DNP2}).

A notorious example is provided by 
the matrix algebra of $2\times 2$ complex Hermitean matrices $H(2, \C)$, whose basis can be realized by the well-known Pauli matrices, and which plays a key role in the description of the Kepler model \cite{Popov}.
The matrix multiplication $\cdot$, obeying 
$
\sigma_i \cdot \sigma_j = \delta_{ij} \sigma_0 + i \epsilon_{ijk} \sigma_k  
 \ ,
$
with respect to the Pauli basis, a generic element of $H(2, \C)$ generally has components lying in a Lorentz vector module in $3+1$ space-time dimensions:
$$X=X^\mu \sigma_\mu= \left(\ba{ccc} X^0 + X^3&\quad& X^1+iX^2\\
 X^1-i X^2&& X^0 - X^3 \ea\right) \ , \qquad \mu=0,1,2,3 \ .$$ 
Moreover, when endowed with a  commutative (Jordan) product  $\circ$ via (half of) the  matrix anticommutator,
$\sigma_\mu \circ \sigma_\nu = \delta_{\mu, \nu} =\demi\{\sigma_\mu, \sigma_\nu \} \ ,$
$H(2, \C)$ acquires the structure of a simple, rank 2 Jordan algebra, namely $\JJ_2^\C= ( H(2, \C), \circ)$.
Also, the norm of an element $X$ of the algebra $\JJ^\C_2$ matches the determinant of its matrix realization given above :
$
\det X 
= (X^0)^2- (X^1)^2-(X^2)^2-(X^3)^2,
$
which coincides with the  Lorentz invariant. All in all, the real  vector  space $H(2, \C)$ becomes
the simple rank 2 Jordan algebra $\JJ_2^\C= ( H(2, \C), \circ)$, and in turn one obtains the isomorphism between $(\JJ^\C_2, \det)$ and the Minkowski space $\R^{1,3}$
with metric
 $g_{\mu \nu} =diag(+,-,-,-)$.

On the other hand, the Pauli matrices 
$\sigma_\mu$ are also Dirac gamma matrices in 
an associative Clifford algebra, and this implies the existence of the (accidental) Jordan algebraic isomorphisms  :
$
\JJ_2^\C \cong \JJ Spin_3,
$
where $\JJ Spin_3$ is Lorentzian spin factor in four dimensions, which is  rank 2 simple Jordan algebra.

To the rank 2, simple Jordan algebra $\JJ_2^\C$,  the TKK correspondence associates a chain of Lie algebraic embeddings, given by (\ref{Tkk}), which specifies here as

\[
\ba{ccccccccc}
\mathfrak{der}(\JJ^{\C}_2) &\cong &
\mathfrak{su}(2) &\cong& \mathfrak{so}(3) &\cong & \mathfrak{der}(\JJ Spin_3)\\
\cap & \\
\mathfrak{str}_0(\JJ^{\C}_2) &\cong &
\mathfrak{sl}(2,\C)&\cong&\mathfrak{so}(1,3)&\cong & \mathfrak{str}_0(\JJ Spin_3) \\
\cap &  \\
\mathfrak{co}(\JJ^{\C}_2) &\cong &
\mathfrak{su}(2,2)&\cong&\mathfrak{so}(2,4)&\cong & \mathfrak{co}(\JJ Spin_3) 
\ea
\]

This chain of algebras induces the Lie group inclusions    
{
\[
\ba{rlc}
Aut(\JJ^{\C}_2) 
&  Spin(3) &  $Rotations $M_{ij}   \\
Str_0(\JJ^{\C}_2) 
&  Spin(1,3)  &  $ Lorentz group $ M_{\mu \nu} \\
Str(\JJ^{\C}_2)
& Spin(1,3) \times \R_+  &  $Lorentz + dilatation $
\{ M_{\mu \nu},  D \} \\
Co(\JJ^{\C}_2) &  Spin(2,4) 
& \hspace{-10pt} $causal space-time automorphisms $ 
\{ P_\mu, K_\mu,M_{\mu \nu}, D \} 
\ea
\]}
%
%
The message of our paper is that Jordan (super)algebras are an elegant  and efficient language when articulating  superconformal symmetries. 

\section{Newton-Hooke duality}
In the  quantum Kepler problem\footnote{For a recent pedagogical review of the Kepler problem, and its quantization, see  \cite{Baez}.},  the wavefunctions of the orbitals transform according to a massless $SO(2,4)$-representation
with helicity $s=0$ \cite{BSW, MT}.
The transitions between the  electron  levels in the hydrogen atom are 
governed by exchanges of photons 
between the electron and the nucleus. The zero mass of the photon, carrier of the electromagnetic interaction, implies the Maxwell equations to be conformally invariant, i.e. to have the Lie algebra $\mathfrak{so}(4,2)$ (and the corresponding Lie group $SO(2,4)$) as symmetries, in the empty space-time interval among the charges and masses in which the photons propagate.
Moreover, in the non-relativistic framework at hand (hydrogen atom), the spin of the electron is neglected, so the photon carries no additional spin.

 A particle possessing both electric and magnetic charge is called a \textit{dyon}.  The charge-dyon system is also referred to as \textit{dyogen atom} \cite{dyogen}, and it stems from the quantization of the  MICZ-Kepler  system \cite{ MIC, Z, MTz} in presence of non-zero magnetic charge $s\neq 0$. On the other hand, the (countably infinite) family of rank 2 Jordan algebras given by the so-called Lorentzian spin factors $\J Spin_D$ in $D+1$ space-time dimensions is suitable for a description of the (MICZ-)Kepler system with an inherent $SO(2,D+1)$ conformal symmetry \cite{Meng, MZ}. In turn, upon  quantization, these  generalize the non-relativistic hydrogen (dyogen) atom   in $\R^D$,  $D\geq 2$ . 

The generalized
 Kunstaanhiemo-Stiefel (KS) transform embeds
 the  hydrogen  (dyogen) atom   into 
the bosonic $s=0$ (fermionic  $s=\demi$ ) sector of a harmonic  oscillator  \cite{Bars,BSW}
\[
\ba{rcl}
$MICZ-Kepler model  in $\R^D \quad &\stackrel{KS}{\leftrightarrow}&  \quad $harmonic oscillator in $\R^{ \bar{D}} \ , \\[4pt]
\mbox{Newton 
}  V_s(r)= -\frac{1}{r} +\frac{s^2}{r^2} 
 \quad &\stackrel{Newton-Hooke}{\leftrightarrow}&  \quad  \mbox{Hooke 
 } U(r) = r^2  \ .
\ea
\]
Consequently, the KS transform is naturally referred to also as the Newton-Hooke duality (in general $D\neq \bar{D}$), 
\cite{DNP,GiP}.

At classical level, a Kepler  orbit is an intersection of the light cone $\det X = g_{\mu \nu}X^\mu X^\nu=0$ with a plane normal  to a given Minkowski  vector \cite{Meng2}. One can obtain three different conic sections,
according to the type of the normal vector (n.v.)  and the 
total energy $E$:

(i)  elliptic orbit (timelike n.v., $E<0$) ,

(ii) hyperbolic orbit  (spacelike n.v., $E>0$) , 

(iii) parabolic (lightlike n.v., $E=0$) .

The  
conformal Lie group  $SO(2,4)$ in four Lorentzian space-time dimensions is a symmetry of the light cone in the Minkowski space $\R^{1,3}$. Remarkably, the action of the  identity component $SO_0(2,4)$ preserves the type of conic section, thus providing a consistent and elegant explanation of the hidden conformal symmetry of  the classical Kepler problem in $\R^3$.

		\section{Landau problem as 2D oscillator }

The  Jordan algebraic inclusion  $\JJ_2^\R \subset \JJ_2^\C$ of the real Pauli matrices into the complex ones defines
a $SO(2,3)$-invariant cone in $\R^{1,2}$. In this case, the (accidental) Jordan algebraic isomorphism $\JJ_2^\R\cong \JJ Spin_2$ holds, and the TKK constuction correspondingly mirrors the Lie group isomorphism
\beq
Sp(4, \R)/\Z_2 \cong SO_{0}(2,3) \ .
\eeq

Let us consider the simple example of the Newton-Hooke duality with $D=\bar{D}=2$. A harmonic oscillator in the plane $\R^2$ is in fact equivalent
to the quantum harmonic motion of an electron in a transversal magnetic field, the so called  {\it  Landau problem}, whose spectrum generating symmetry is provided by the group $Sp(4, \R)$, which indeed acts on the so-called \textit{Landau levels} \cite{DNP}.

In complete analogy with the Kepler model in $\R^3$,  the non-relativistic hydrogen atom in $\R^2$ is defined as the massless irreducible unitary $SO(2,3)$-representation of helicity $s=0$. On the other hand, the dyogen atom in $\R^2$ is the massless irreducible unitary $SO(2,3)$-representation of helicity $s=\demi$, and it describes the interaction of a magnetic vortex with a charge \cite{NTT,TerANersessian}.

The  Fock space of the Landau levels can be constructed in terms of two decoupled Heisenberg algebras, generated by the energy creation and annihilation modes $a^{\pm}$ and by the magnetic translations $b^{\pm}$. The Landau Hamiltonian and the angular momentum operators read
\[
H= \omega \left(a^+ a^- + \demi\right) \ , \qquad L_z= \demi \left(b^+b^- - a^+a^- \right),  \qquad [L_z, H]=0  \ ,
\]
and they both belong to the Cartan subalgebra of 
the  conformal algebra $\mathfrak{sp}(4,\R)\cong\mathfrak{so}(2,3)$, which is indeed generated by  all bilinear polynomials of  $a^\pm$ and $b^\pm$.
Together with the parity operator $\Pi$, the operators $H$ and $L_z$ form 
a set of mutually commuting operators
$\{H, L_z, \Pi \}$, which can therefore be  simultaneously  diagonalized :
\[
H \psi_{n,m}=
 \omega \left(n+\demi\right) \psi_{n,m} \ , \quad L_z \psi_{n,m}=
m \psi_{n,m} \ , \quad
\Pi \psi_{n,m}= (-1)^m \psi_{n,m} \ .
\]
Here the natural number $n\geq 0$ labels the $n$-th Landau level, whereas $m\geq -n$ is the magnetic quantum number\footnote{We adopt the system of units with  $\hbar=c=e=1$ throughout the text.}; the constant $\omega$ is the cyclotronic frequency.

\

A suitable linear combination of $H$ and $L_z$ defines the harmonic oscillator Hamiltonian, which underlies the so-called \textit{planar Landau problem}.
\[
H_{osc} =H+ \frac{\omega}{2} L_z \ , \qquad 
H_{osc}  \psi_{n,m}= 
E \psi_{n,m} \ , \quad 
 L_z \psi_{n,m}= m\psi_{n,m} \ .
\]

Thus, $H_{osc}$ can be diagonalized in the same Fock space  as the Landau Hamiltonian $H$ and the angular momentum $L_z$.
Notice that $H_{osc}$ is however infinitely degenerate in $L_z$. Its  eigenvalue $E=\omega \left(n+\frac{m}{2}+ \demi\right)$ is finitely degenerated, yielding the so-called conformal energy, $E$ is finitely degenerate.
The helicity $s$ of an irreducible $SO(2,3)$-representation is the intrinsic angular momentum (spin),  the minimal value of the angular momentum $L_z$ of the (pseudo)vacuum state $\psi_{0,2s}$. 

Dirac's remarkable representation of the{ anti-}de Sitter (AdS) group $SO(2,3)$ \cite{Dirac}
was shown to be isomorphic to the  Fock space ${\bf Dirac }$ of the Landau problem \cite{DNP, DNP2},
\beq
\label{hilb}
{\bf Dirac } = \bigoplus_{n\geq 0} {\bf Dirac }_n=
\bigoplus_{n\geq 0}\,\, \bigoplus_{m\geq -n}  \psi_{n,m}(z , \bar{z}) \ .
\eeq
In this framework, the parity operator $\Pi\in \Z_2 $ 
singles out two orbits under the action of the the spectrum generating group
 $SO(2,3)$-symmetry.

Moreover, the  Levi-Civita transform $w=z^2$ provides an implementation of the Newton-Hooke correspondence, which relates
the  odd $\mathbf{Di}$ resp.  even $\mathbf{Rac}$ Fock subspaces  of the  Landau levels of the harmonic oscillator  Hilbert space onto the dyogen resp. hydrogen  atom in the plane $\R^2$.
\[
  \left\{\mbox{Dyogen atom} \right\} \oplus \left\{\mbox{Hydrogen atom} \right\} \quad
	\stackrel{Newton-Hooke}{\longleftrightarrow}  \quad \mathbf{Di}
\oplus 
\mathbf{Rac}  \ .
\]
On Newton's side of the correspondence, the spectrum generating group $SO(2,3)$ acts on the regularized Coulomb-Kepler quantum systems
(with or without  magnetic charge).
On Hooke's side, the double covering of the (identity connected component of the)
 AdS group $SO(2,3)$
has two disjoint orbits: namely, the odd resp. even states, built upon the fermionic resp. bosonic ground state\footnote{Alternatively, one can make the choice $\psi_{-2s,2s}$ for the ground states.},
$\psi_{0,2s}$  with parity $(-1)^{m}=(-1)^{2s}$. 
 
The unitary massless representations of
$SO(2,3)$ are denoted by $D(E_0, s)$; in fact, being such a Lie group of rank 2, they are classified by the eigenvalues of the only two independent Casimirs, namely the so-called minimal {\it conformal energy} $E_0$  and the {\it helicity} $s$.
Consequently, as discovered by Flato and Fronsdal in their pioneeristic paper {\it "Massless particles, conformal
group, and de Sitter universe" }\cite{FF} (whose we adopt here the notation), the Fock 
 space of the Landau levels splits into two \textit{singleton} $Sp(4, \R)$-representations : the planar dyogen atom (i.e., the quantum system of an electron and a charged magnetic vortex) has Hilbert space $\mathbf{Di}=D(E_0=1, s=\demi) $,  whereas the planar hydrogen atom has Hilbert space $D(E_0=\demi, s=0)$. Namely :
\[
{\bf Dirac } = {\bf Dirac }_{\bar{1}} \oplus  {\bf Dirac }_{\bar{0}} = \mathbf{Di}
\oplus 
\mathbf{Rac}
= D\left(1, \demi\right) \oplus D\left(\demi, 0\right) \ . 
\]

A consistent superization of the AdS group, yielding to a consistent realization of the whole Fock space of Landau levels in terms of super-oscillators, can be achieved as follows : if one adds a fermionic oscillator with modes $\psi$ and $\psi^\dagger$, and moreover
\[
a^\dagger = a^+ \psi \ , \quad  a = a^- \psi^{\dagger} \ , \quad b^\dagger = b^+ \psi  \ ,\quad   b = b^- \psi^{\dagger} \ , \qquad
\{ \psi^\dagger , \psi \}=1 \ ,
\]
then  
the full Fock space ${\bf Dirac } $ is generated by $a^\dagger$ and $b^\dagger$ only, acting on the ground state $ | 0 \rangle := \psi_{0,0}  $
and it sits into a single irreducible \textit{supersingleton} representation
of the Lie superalgebra $\mathfrak{osp}(1|4, \R)$, which splits as 
$$\mathfrak{osp}(4, \R)_{\bar{0}}\oplus \mathfrak{osp}(4, \R)_{\bar 1}   = \mathfrak{sp}(4, \R) \oplus \left( \ba{c} a^+   \\
b^+ \ea\right) \psi_{0,0}
 = \mathfrak{sp}(4, \R) \oplus \left( \ba{c} \psi_{-1,1}  \\
\psi_{0,1} \ea\right)
$$
with bosonic 
 $\psi_{0,0} \in \mathbf{Rac}$  and fermionic  
 $\psi_{0,1},\psi_{-1,1} \in \mathbf{Di}$ ground states. One may thus conclude that the Lie supergroup
$OSp(1|4, \R)$ is the dynamical supersymmetry of the planar Landau problem, see \cite{DNP2}.
The even generator, 
{
the parity}   $\Pi= [\psi^\dagger, \psi]$ extends $OSp(1|4, \R)$ to $OSp(2|4, \R)${
\footnote{ 
Alexander Ganchev and Chavdar Palev \cite{GP}
showed that the algebra closed by $n$ bosonic oscillators is the
parabosonic algebra $OSp(1|2n, \R)$. Its extension with the parity operator $\Pi$ is shown to be isomorphic to  the algebra $OSp(2|2n, \R)$ in \cite{NJ}.}} which is the 
superconformal symmetry of the
supersymmetrized  Landau model with a term $B\sigma_3$, see (\ref{1D}) below, accounting for the interaction of 
the electron  spin with the magnetic field $B$. 

An inspection of the TKK construction for the Jordan superalgebras yields to the following chain of embeddings of Lie superalgebras for the exceptional Kac-Jordan superalgebra $\JJ F$ (of superdimension $(6|4)$) (see e.g. \cite{BC}) :
\[
\ba{ccccc}
\mathfrak{der}(\JJ F)  &\subset&  \mathfrak{str}_0(\JJ F) &\subset& \mathfrak{co}(\JJ F) \ , \\[4pt]
\mathfrak{osp}(1|2,\R )^{\oplus 2}  &\subset&   \mathfrak{osp}(2|4,\R)  &\subset&  \mathfrak{f}(4)  \ .
\ea
\]
Thus, the reduced structure Lie superalgebra $\mathfrak{osp}(2|4,\R)$ of $\JJ F$ matches the superconformal Lie algebra of the superized Landau problem with spin. 
We here put forward the conjecture that there exists a superized of the Landau problem
which has dynamical supersymmetry (i.e. superconformal Lie algebra) given by the exceptional Lie superalgebra $\mathfrak{f}(4)$.

\section{Landau problem as 1D oscillator}
 

The Landau Hamiltonian with spin interaction  can consistently be rewritten by the ladder operators $a$ and $a^\dagger$ of only one Heisenberg superalgebra (see  \cite{Hasebe, Smilga}):
\beq
\tilde{H} = \frac{\omega}{2} \{ a^+, a^-\} - \frac{\omega}{2} [\psi^\dagger  , \psi] = \frac{\omega}{2} \{ a^\dagger, a\},
\label{1D}
\eeq
whereas the  magnetic translation operators $b^\pm$ are energy zero-modes  and explain 
 the infinite degeneracy of the Landau levels.
If one chooses to disregard the quantum number $m$, which accounts for the degeneracy of the Landau levels,
then the interesting option arises that the Landau problem actually can be reformulated in terms of one 1D (super)oscillator only. In fact, this is the common approach 
in  literature on the  Landau problem supersymmetric extensions, 
 see e.g. \cite{DHE, Plyushchay, EI}. 
{
It is worth noting that the
superconformal symmetry is inherent   to both the Landau levels and  the Regge trajectories  in the spectra of
mesons and baryons. It has a reincarnation in the
Light-Front QCD holography approach
of Stanley Brodsky 
(see \cite{Brodsky} and references therein).}



{\bf Superconformal symmetries from Kaplansky superalgebra $\JJ\R^{1|2}$}

\
In this note, we aim at putting forward the proposal to apply the (superized version of the) TKK correspondence to the Kaplansky 
Jordan superalgebra $\JJ\R^{1|2}$ in order to obtain the superconformal symmetries of  the 1D super-oscillator (with various degrees of super-extensions, ranging from $N=1$ to $N=4$).

The Kaplansky superalgebra $\JJ\R^{1|2}$ is a Jordan superalgebra of superdimension ${(1|2)}$, and it can be regarded as the (minimal) superization (i.e., the supersymmetric extension) of the simple Jordan algebra given by the real numbers $\R$. Such a minimal superization introduced two two Grasmannian variables
 $\theta^{\alpha} =(\theta^1,\theta^2) $,
 cf. the classification of Victor Kac \cite{V}.
The Kaplansky superalgebra $\JJ\R^{1|2}$ admits a homogeneous basis 
 $e_{A}=(e_{0},e_{\alpha})$ 
 where $e_{0}$ span the even (bosonic) time coordinate and $(e_{1},e_{2})$ span the odd (fermionic) 'times'. An element $\j \in \JJ\R^{1|2}$  can thus be written as
\beq
\j=
e_{0} t+ e_{\alpha} \theta^{\alpha} \ , \qquad \alpha=1,2 \ ,
\eeq
where $
t\in\mathbb{R}$ is can be thought as the real, actually physical 
time coordinate, and $\theta^{\alpha}$ are 'Grassmann 
time' coordinates.
 The composition rules of the (super)Jordan product $\circ$ in $\JJ\R^{1|2}$ are given by
\beq
\ba{cclrclrcl}
e_{0}\circ e_{0}&=&e_{0} \ ,&e_{0}\circ e_{1}&=&\demi e_{1} \ , & e_{1}\circ e_{1}&=&0 \ ,  \\[4pt]
e_{1}\circ e_{2}&=&e_{0} \ , & e_{0}\circ  e_{2}&=&\demi e_{2} \ , &     e_{2}\circ e_{2}&=&0 \ .
\ea
\label{K3}
\eeq

The results of the application of the (possibly extended) super TKK correspondence to the Kaplansky superalgebra $\JJ\R^{1|2}$, with various supersymmetric extensions of the Poincar\'e half-plane $H^1\cong\R \times \R_{\geq 0} \cong D^1$:


\begin{center}
\begin{tabular}{|c|c|c|c|c|}
\hline
$\JJ \R^{1|2}$&$N=0$&$N=1$ & $N=2$ & $N=4$ \\
\hline
 TKK& $\mathfrak{co}(\JJ \R)$ & $\mathfrak{der}(\JJ\R^{1|2})$ & $\mathfrak{str}(\JJ\R^{1|2})$ & $\mathfrak{co}(\JJ\R^{1|2})$
\\ \hline
 $\mathfrak{co}(\JJ\R^{1|2})$  &$\mathfrak{sp}(2,\R)\cong \mathfrak{su}(1,1) $& $\mathfrak{osp}(1|2)$
& $\mathfrak{osp}(2|2)\cong \mathfrak{su}(1,1|1)$ & $\mathfrak{su}(1,1|2)$ \\ \hline
Model &dAFF \cite{DFF}&$\ba{c} $ Landau$ \\
$no spin$ \ea$& 
$\ba{c}
 $Landau$
\\ $with spin$
\ea$ & $AdS_2\times S^2$
\\ \hline
$\ba{c}
$Bounded$\\$Domain$
\ea$ &
$H^{1}$
& $H^{1|1}$&  $H^{1|2}$ &$H^{1|4}$ \\
\hline
\end{tabular}
\end{center}
Consistently, the supersymmetries of the 1D oscillator arise via a reduction of the 2D oscillator system; indeed, the (super)Jordan algebraic inclusion $\JJ \R^{(1|2)}\subset {\JJ F}$ implies
\beq
\left( \mathfrak{der}(\JJ\R^{1|2}) \cong \mathfrak{osp}(1|2) \right) \subset  
\left(  \mathfrak{ der}({\JJ F}) \cong \mathfrak{osp}(1|2)^{\oplus 2} \right).
\eeq
As a further remark on the table above, we observe the following curious fact. While $\mathfrak{osp}(1|2)$ and $\mathfrak{osp}(2|2)$ can rightfully be regarded as the $N=1$ resp. $N=2$ superconformal Lie algebra in one (timelike) dimension, the conformal algebra of the Kaplansky superalgebra, $\mathfrak{co}(\JJ\R^{1|2})=\mathfrak{su}(1,1|2)$, is not a space-time superconformal Lie algebra itself, but it rather is a (co-rank one) subalgebra of the actual $N=4$ superconformal Lie algebra in one dimension, $\mathfrak{osp}(4|2)$ :
\beq
\mathfrak{co}(\JJ\R^{1|2})=\mathfrak{su}(1,1|2)\subset \mathfrak{osp}(4|2).
\eeq

\

{\bf Kaplansky derivations $\mathfrak{der}(\JJ\R^{1|2})$ as $N=1$ superconformal Lie algebra of 1D oscillator}
\

The odd generators of the $N=1$ superconformal algebra $\mathfrak{osp}(1|2)$ of the 1D oscillator are the supercharges $Q$ and $S$,  they transform as  a real (i.e., Majorana) spinor 
\[
Q_ \alpha= (Q_1, Q_2) = ( Q, S) \ , \qquad Q^\beta= \epsilon^{\beta \alpha}  Q_\alpha  \ , \qquad Q_\alpha^\dagger= Q_\alpha  \ , 
\]
under the action of the even subalgebra  $\mathfrak{osp}(1|2)_{\bar{0}}= \mathfrak{sp}(2,\R)$ 
with generators
\[
T_{\alpha \beta} = \demi \{ Q_\alpha, Q_\beta \}  \  = \left( \ba{cc}  H & D \\ D & K\ea \right) \ .
\]
Thus, $Q_\alpha$ is a Majorana 2-dimensional spinor in $2+1$ space-time dimensions, and its components  $Q$ and $S$ stem from the commutators of Jordan multiplication operators $L_A (x)= e_A \circ x$ on a basis $e_A$ (cf. (\ref{K3})),
	\[ Q_\alpha \sim D_{0\alpha}=[L_0, L_\alpha] \ , \qquad \alpha=1,2.
	\]
Thus, $Q_\alpha$ belongs to   $\mathfrak{der}(\JJ\R^{1|2})$ too.

By adopting the succinct notation for $\mathfrak{osp}(1|2)$ used in the pedagogical review of superconformal mechanics \cite{FIL}, the (graded) commutation relations\footnote{With respect to the usual relations, let us notice the sign difference, due to our convention $\epsilon_{12}=\epsilon^{21}=-1$.} of $\mathfrak{osp}(1|2)$ reads
\beq
[T_{\alpha \beta}, T_{\gamma \delta}] = - i ( \epsilon_{\alpha \gamma} T_{\beta \delta} - \epsilon_{\beta \delta} T_{\alpha \gamma}) \ , \qquad
[T_{\alpha \beta}, Q_\gamma] = \frac{i}{2}( \epsilon_{\gamma \alpha} Q_\beta + \epsilon_{\gamma \beta} Q_\alpha) .
\eeq
\textit{In extenso}, such relations read
\beq
\ba{lll}
[D, H] = - i H  \ , & [H, K] = 2i D  \ , &  [D, K ] = i K \ , \\[4pt]
[D, Q] = -\frac{i}{2} Q  \ , & [H,Q]=0 \ ,       & [D, S] = \frac{i}{2} S \ ,\\[4pt] 
  [H,S]=  i Q \ , & [S,K]=0 \ ,  & [K, Q] = -  iS \ , \\ [4pt]
\{Q,Q\} = 2H \ , &  \{Q, S\} = 2 D   \ , & \{S,S\} = 2K \ .
\ea
\label{osp(1|2)}
\eeq
It is here worth remarking that the dilatation $D$ is a grading operator, yielding the following 5-grading of the Lie superderivation algebra $\mathfrak{osp}(1|2)$ (here characterized as the $N=1$ superconformal algebra of the 1D oscillator):
\[
\d:= \mathfrak{der}(\JJ\R^{1|2}) \cong \underbrace{\d_{-1}}_{H} \oplus  \overbrace{\d_{-\demi}}^{Q} \oplus \underbrace{\d_{0}}_{D}
 \oplus \overbrace{\d_{+\demi}}^{S} \oplus \underbrace{\d_{+1}}_K \ .\]
The even subalgebra of the $N=1$ superconformal Lie algebra $\mathfrak{osp}(1|2)$ is $3$-graded, and, as expected, it coincides with the conformal Lie algebra of the even part of the Kaplansky superalgebra $\JJ\R^{1|2}$, namely with the conformal symmetry of the time $\R$, regarded as simple Jordan algebra  :
\[
\mathfrak{der}(\JJ\R^{1|2})_{\bar{0}} \cong \underbrace{\d_{-1}}_{H} 
\oplus \underbrace{\d_{0}}_{D}
\oplus \underbrace{\d_{+1}}_K \cong \mathfrak{co} (\JJ \R) \cong \mathfrak{sp}(2, \R)\ .
\]
As already pointed out, this is not surprising, as the superderivations  $\mathfrak{der}(\JJ\R^{1|2})$ are a supersymmetric extension  of $\mathfrak{sl}(2, \R) \cong \mathfrak{sp}(2, \R)$.

The (unique, quadratic) Casimir of the superalgebra $\mathfrak{osp}(1|2)$ reads
\[
C_2 = \demi T_{\alpha \beta} T^{\alpha \beta} -
\frac{i}{2} Q_\alpha Q^\alpha = \demi \{ K, H \} -D^2 -
\frac{i}{2} [ Q, S].
\]

{\bf  5-grading of $\mathfrak{der}(\JJ\R^{1|2})$ by the compact generator $T_0$}
\

The conformal energy $T_0$ is another operator  that can play the role of $\mathfrak{der}(\k)$-grading. 
The new odd  generators diagonalizing $T_0$ are dimensionless combinations of the supercharges $Q$ and $S$
\[
\Gamma_\pm = \frac{1}{\sqrt{2}} \left( \sqrt{m} S \pm i \frac{Q}{\sqrt{m}} \right)\ .
\]
The  superconformal algerba $\mathfrak{osp}(1|2)$ generated by the set $\{ T_0, T_\pm , \Gamma_\pm \}$ has  the following nonvanishing relations 
\[
\ba{ccccccr}
[T_0 , T_\pm ] &= &\pm T_\pm  \ , &\qquad  &[T_+, T_- ]& =& - 2 T_0 \ , \\[4pt]

[T_0 , \Gamma_\pm ] &= &\pm \demi \Gamma_\pm  \ , &\qquad  &\{\Gamma_+, \Gamma_- \}& =&  2 T_0 \ , \\[4pt]
[T_\mp, \Gamma_\pm ]& =&  \pm \Gamma_\mp
  \ , &\qquad  & \{\Gamma_\pm , \Gamma_\pm \} &= &\pm 2 T_\pm \ . 
	\ea
\]

We got an alternative form $\tilde {\d}$ of $\mathfrak{osp}(1|2)$ with a  $T_0$-grading 
\[
\tilde {\d} \cong \mathfrak{der}(\k) \cong \underbrace{\tilde {\d}_{-1}}_{T_-} \oplus  \overbrace{\tilde {\d}_{-\demi}}^{\Gamma_-} \oplus \underbrace{\tilde {\d}_{0}}_{T_0}
 \oplus \overbrace{\tilde {\d}_{+\demi}}^{\Gamma_+} \oplus \underbrace{\tilde {\d}_{+1}}_{T_+} \ .
\]
In this basis, the Casimir of the superalgebra reads
\[
C_2 = \demi T_{\alpha \beta} T^{\alpha \beta} - 
\frac{i}{2} Q_\alpha Q^\alpha = T_0^2 - \demi \{ T_+, T_- \} - \demi [\Gamma_+ , \Gamma_-].
\]

{\bf Parabosonic algebra.}An alternative to the canonical quantization coined paraquantization was proposed by H. S. Green. 
The parabosonic algebra with $n$-modes turn out to be isomorphic to the supersymmetric algebra $\mathfrak{osp}(1| 2n)$ \cite{GP}
\[
[\{a^+_i, a^-_j\}, a^\pm_k ] = 2 \delta_{jk} a^\pm_i \ , \qquad [\{a^\pm_i, a^\pm_j\}, a^\pm_k ] =0 \ .
\]
The odd generators $a^\pm_i$ yield an oscillator basis for the superalgebra of type  $B_{(0|n)}$ in the Kac tables.
If one takes only one paraboson $n=1$ one ends up with the algebra $\mathfrak{osp}(1| 2)$  whose relations imply

\[
[ \{\Gamma_+, \Gamma_-\}, \Gamma_\pm ] = 2  \Gamma_\pm  \ , \qquad [\{\Gamma_\pm, \Gamma_\pm\}, \Gamma_\pm ] = 0 \ .
\]

We conclude that the isomorphism between the superconformal algebra  $\mathfrak{osp}(1| 2)$  and the paraboson algebra
is given by
\[
\Gamma_\pm = \frac{a^\pm}{\sqrt{2}} \ , \qquad  T_\pm = \demi (a^\pm)^2 \ , \qquad T_0= \frac{1}{4} \{a^+, a^- \}  \ .
\]
At the end of the day all the $\mathfrak{osp}(1|2)$-generators
are quadratic and linear monomials of $a^+$ and $a^-$ within
one Heisenberg algebra, see also \cite{Waldron, Henkel}.
{
The extension with the parity operator $\Pi=[\psi^\dagger, \psi]$
introduces the spin in the Landau hamiltonian, eq. (\ref{1D}) having extended  supersymmetry
$\mathfrak{osp}(2|2)$ \cite{NJ}.}

\section{Outlook and perspectives}

The TKK correspondence is a natural algebraic framework for interpreting the dynamical symmetries of the charged spin-0 and  spin-$\demi$ particles in a magnetic background, linking supersymmetric quantum mechanics to the structure of Jordan superalgebras, namely, in our treatment, to the Kaplansky superalgebra $\JJ\R^{1|2}$ and to the exceptional Jordan superalgebra $\JJ F$.

The orthosymplectic  Lie superalgebra $\mathfrak{osp}(1|2)$ is the derivation Lie algebra of the Kaplansky superalgebra  $\JJ\R^{1|2}$,
$$\mathfrak{der}( \JJ\R^{1|2} )\cong \mathfrak{osp}(1|2) = \mathfrak{osp}(1|2)_{\bar{0}} \oplus \mathfrak{osp}(1|2)_{\bar{1}}= \mathfrak{sp}(2)\oplus Q_\alpha \ , $$
 and it can also be conceived as the minimally supersymmetric ($N=1$) superconformal Lie algebra of the 1D oscillator \cite{Waldron}. In fact, it is spanned by
 the quadratic and linear polynomials of  1D oscillators $[a^-, a^+]=1$, e.g., by  the energy creation-annihilation operators $a^{\pm}$
 or magnetic translations $b^{\pm}$. 

The  time reparametrization  
group  $SL(2, \R)$ in conformal mechanics,
\[
t \mapsto t^\prime = \frac{\alpha t + \beta}{\gamma t +\delta} = e^{iaH} e^{ibD} e^{icK}\quad \mbox{where} \quad 
SL(2, \R)/ \Z_2 \cong  SO_0(1,2)
\]
is a symmetry of de Alfaro-Fubini-Furlan (dAFF) model \cite{DFF}, whose Hamiltonian reads
\[
H = \frac{p^2}{2m} + \frac{g}{2x^2} \ .
\]
The  field $x(t)$ is defined in $0+1$ space-time of conformal dimension $\Delta=\demi$. The Hamiltonian $H$, the time dilatation $D$
and the special conformal generator $K$ span the algebra
$\mathfrak{co}({\JJ \R})$ which we will
call (after Barut)  {\it radial algebra} \cite{BSW, IKL}  
$$ \mathfrak{co}({\JJ \R}) \cong   \mathfrak{sl}(2,\R) \cong\mathfrak{sp}(2, \R)\cong \mathfrak{so}(1,2) \cong \mathfrak{su}(1,1)\ .$$
 Therefore, the conformal TKK algebra $\mathfrak{co}({\JJ \R})$
of the time (1D) Jordan algebra ${\JJ \R}$ 
 contains  an Hamiltonian along with operators lowering/raising
 the energy.
 
In general, the radial algebra $\mathfrak{co}({\JJ \R})$ arises in conformal mechanics
as a  part of the larger spectrum generating algebra $\mathfrak{co}(\JJ^\prime)$, since one always has time (energy) inclusion  ${\JJ \R} \subset \JJ^\prime$ 
into a larger Jordan algebra of observables $\JJ^\prime$. 
Separation of variables singles out the evolution along the radial degree of freedom via the Schr\"odinger equation.
Examples of {\it radial algebras } 
include :

\textbf{i}) Barut's  
 $\mathfrak{so}(1,2)\subset \mathfrak{so}(2,4)$ for the hydrogen and dyogen atom \cite{BSW},

\textbf{ii}) Jackiw's dynamical symmetry
$\mathfrak{so}(1,2)$ of the magnetic monopole 
\cite{RJ}.



The minimally supersymmetric extension of the radial algebra
$\mathfrak{co }(\JJ \R )\subset \mathfrak{der } (\JJ\R^{1|2})$
is naturally referred to as a $N=1$ { \it radial superalgebra} $\mathfrak{der } (\JJ\R^{1|2}) \cong \mathfrak{osp}(1|2)$.
  The Jordan algebraic inclusion 
  implies the Lie algebra inclusions via the TKK correspondence :
\beq
\JJ \R \subset \JJ\R^{1|2} \qquad \Rightarrow \qquad
\underbrace{\mathfrak{sp}(2, \R)}_{N=0}\subset
\underbrace{\mathfrak{osp}(1|2)}_{N=1}\subset
\underbrace{\mathfrak{osp}(2|2)}_{N=2}
\subset \underbrace{\mathfrak{su}(1,1|2)}_{N=4} \ .
\eeq
Hence, the TKK construction for the  Kaplansky superalgebra $\JJ\R^{1|2}$ 
implies 
a hierarchy of superextensions of the radial algebra $\mathfrak{co} (\JJ\R)$, given by the
$N=2^k$(-extended)  { \it radial superalgebras},
namely $\mathfrak{der } (\JJ\R^{1|2})$, $\mathfrak{str } (\JJ\R^{1|2})$
and $\mathfrak{co } (\JJ\R^{1|2})$ for $k=0,1,2$, respectively.

We make the observation that the  superextension $SU(2|1)$ of  the Landau problem considered by Evgeny Ivanov \cite{EI} and collaborators 
matches with our  $N=2$ { \it radial superalgebra} $\mathfrak{str } (\JJ\R^{1|2})$ \cite{AV, FR}. Other radial superalgebras  in the context of the  $AdS_2/CFT_1$ correspondence, Haldane sphere and near-horizon black hole dynamics will be given elsewhere.

In a future work, we are also planning to combine ideas from Jordan superalgebras, Itzhak Bars' 2 time physics and the superfield approach  \cite{Bars1, Bars,FIL,IKL2, Henkel2}.

\begin{acknowledgement} T.P. would like to
thank  Michel Dubois-Violette, Ivailo Mladenov, Mustafa Mullahasanoglu,  Petko Nikolov, Philippe Nounahon, Ramakrishna Ramaswamy for their encouraging interest in this work and many enlightening discussions. T.P. also received partial support from the Bulgarian NSF grant KP-06-N68/3.
\end{acknowledgement}

\end{document}